\documentclass{aa}
\usepackage{graphicx}
\usepackage{txfonts}
\usepackage{natbib}
\usepackage{hyperref}
\hypersetup{colorlinks=true, citecolor=blue}
\usepackage{graphicx}
\usepackage{txfonts}
\usepackage{rotating}
\bibpunct{(}{)}{;}{a}{}{,} 

\begin{document} 
   \title{Impact of star formation history on the measurement of star formation rates}
   \author{M. Boquien\inst{1} \and V. Buat\inst{2} \and V. Perret\inst{3}}
   \institute{Institute of Astronomy, University of Cambridge, Madingley Road, Cambridge, CB3 0HA, United Kingdom \email{mboquien@ast.cam.ac.uk}
              \and Aix Marseille Universit\'e, CNRS, LAM (Laboratoire d'Astrophysique de Marseille) UMR 7326, 13388, Marseille, France
              \and Institute for Theoretical Physics, University of Zurich, CH-8057 Zürich, Switzerland
   }

   \date{}

  \abstract
   {Measuring star formation across the Universe is key to constrain models of galaxy formation and evolution. Yet, determining the SFR (star formation rate) of galaxies remains a challenge.}
   {In this paper we investigate in isolation the impact of a variable star formation history on the measurement of the SFR.}
   {We combine 23 state--of--the--art hydrodynamical simulations of $1<z<2$ galaxies on the main sequence with the \textsc{cigale} spectral energy distribution modelling code. This allows us to generate synthetic spectra every 1~Myr for each simulation, taking the stellar populations and the nebular emission into account. Using these spectra, we estimate the SFR from classical estimators which we compare with the true SFR we know from the simulations.}
   {We find that except for the Lyman continuum, classical SFR estimators calibrated over 100~Myr overestimate the SFR from $\sim25$\% in the FUV band to $\sim65$\% in the U band. Such biases are due 1) to the contribution of stars living longer than 100~Myr, and 2) to variations of the SFR on timescales longer than a few tens of Myr. Rapid variations of the SFR increase the uncertainty on the determination of the instantaneous SFR but have no long term effect.}
   {The discrepancies between the true and estimated SFR may explain at least part of the tension between the integral of the star formation rate density and the stellar mass density at a given redshift. To reduce possible biases, we suggest to use SFR estimators calibrated over 1~Gyr rather than the usually adopted 100~Myr timescales.}

   \maketitle

\section{Introduction\label{sec:intro}}

The SFR (star formation rate) is a key parameter to understand galaxy formation and evolution. If it plays such a central role, this is due to its fundamental relation with the stellar mass and the gas reservoir on which our understanding of galaxy formation and evolution is based. Yet, despite its considerable importance, measuring star formation accurately remains a challenging task \citep[see][for a recent review]{kennicutt2012a}.

One of the key assumptions to measure the SFR of a galaxy is its SFH (star formation history). Most classical estimators are based on the assumption of a constant SFR over a period of 100~Myr \citep[e.g.][]{kennicutt1998a}. If this assumption may seem reasonable for low redshift spiral galaxies evolving secularly, it is unlikely to hold true for interacting systems or at higher redshift where the SFR necessarily varies on timescales that can be similar or shorter than 100~Myr. With the increasing availability of observations spanning a broad range of wavelengths, one possibility to waive the assumption on the SFH is to carry out multi--wavelengths SED (spectral energy distribution) modelling. The SFH can then be left as a free parameter to compute the SFR. Unfortunately, the SFH is degenerate with other parameters such as the attenuation. For instance a galaxy can exhibit a red UV (ultraviolet) slope because it is dusty and actively forming stars, or because it is not forming stars any more. If multi--wavelengths data partly alleviate this problem when UV and IR (infrared) data are simultaneously available, the SFH still remains poorly constrained.

Even if we cannot uncover the SFH of galaxies in detail, it is nevertheless possible to investigate the impact of short and long term variations of the SFH on the measure of the SFR using numerical simulations of galaxies \citep[e.g.][]{wuyts2009a,wilkins2012b,lanz2014a,simha2014a}. However, simultaneously taking into account both the attenuation and a variable star formation history makes it difficult to isolate the exact impact of the latter on the accuracy of classical SFR estimators. Another potential issue that can be encountered when exploiting numerical simulations is the time resolution of reconstructed SFH. If it is not significantly smaller than the typical time a SFR estimator is sensitive to, then short term variations are in effect smoothed out and their impact cannot be explored and thus remains unknown.

The aim of this article is to explore the impact of long term and short term variability of the SFH on the measurement of the SFR using classical estimators, isolated from any other perturbing effect. In Sect.~\ref{sec:method}, we lay out our method, including how we combine state--of--the--art hydrodynamical simulations of galaxies with a SED modelling code to generate synthetic observations from which we estimate the SFR. We present our results in Sect.~\ref{sec:results} before discussing the best course of action to measure the SFR while taking its variability into account in Sect.~\ref{sec:discussion}. We conclude in Sect.~\ref{sec:conclusion}.

\section{Method\label{sec:method}}

To understand the impact of the variability of the SFH on the measurement of the SFR, our main guideline is to compare estimated SFR from synthetic observations of simulated galaxies to the true SFR we know from the simulations.

In this Section, we first present the set of simulations of high redshift galaxies that provides us with a sample of SFH with a time resolution of only 1~Myr. Then we show how we exploit these SFH to create simulated observations. Finally, we present the set of SFR estimators considered in this study and how they are applied to estimate the SFR from simulated observations.

\subsection{The MIRAGE simulations sample\label{ssec:mirage}}

The use of SFH with plausible variations over all timescales --- from rapid stochastic variations to slow variations over long periods --- is especially important if we want to provide meaningful insight into the impact of the variability of the SFH on the measure of the SFR. This means that we need SFH generated by means of hydrodynamical simulations with a very detailed modelling of baryonic physics and in particular of feedback which can have a dramatic effect in governing the SFR in galaxies. Following these requirements, we rely on the MIRAGE simulations \citep[Merging and Isolated high redshift Adaptive mesh refinement Galaxies,][]{perret2014a}. It comprises a set of 20 idealised mergers simulations sampling four initial disk orientations and five total baryonic masses. The whole merger sample is built using three distinct idealised disk models G1, G2, and G3 that are also simulated in isolation with stellar masses respectively of $\log M_{\star}=9.8$, 10.2, and 10.6. These stellar masses were selected to sample the stellar mass histogramme of a representative spectroscopic sample of $1<z<2$ galaxies, namely the MASSIV sample \citep{contini2012a}. Each model is composed of a stellar disk, a gaseous disk, a stellar bulge, and a dark matter halo. The two disk components are built using an exponential density profile, while the stellar bulge and the dark matter halo are following a common \cite{hernquist1990a} profile. The initial gas fraction of the three disk models is $f_g=0.65$ which is in agreement with the most gas--rich $1<z<2$ population observed \citep{daddi2010b}. These simulations were carried out using the adaptive mesh refinement code RAMSES \citep{teyssier2002a}, allowing to reach a physical resolution of 7~pc in the most refined cells. The modelled gas physics in the simulations includes metal lines cooling, star formation using a local Schmidt law, metallicity advection as well as a recent physically--motivated implementation of stellar feedback \citep{renaud2013a} that encompasses OB--type stars radiative pressure, photo-ionization heating and supernovae. The SFH of each simulation has been obtained by computing a mass--weighted histogramme of the age of the stars particles using a bin of 1~Myr over about 800~Myr.

Because of the high gas fraction of the galaxy models, Jeans instabilities arise in the first dynamical times of the simulations and trigger very high SFR, more than 2$\sigma$ away from the normal star formation regime of the simulations. This transitory relaxation phase that occurs within the first 250~Myr is excluded from our analysis to avoid spurious effects due to its non--representativeness.

While the physical properties of the three disk models of the MIRAGE sample are defined to be comparable to the properties of galaxies observed in the redshift range $1<z<2$, the star formation implementation is not specifically designed to perfectly match the normal SFE (star formation efficiency) of the observations. Indeed, MIRAGE simulations display a SFE slightly under the \cite{daddi2010a} relation, within 1$\sigma$. Nevertheless, this feature is not really relevant for this analysis since we are interested in the relative SFR fluctuations more than the absolute value. Moreover, the merger simulations do not display any SFE enhancement which is interpreted as a saturation mechanism of the star formation in \cite{perret2014a}. The remarkable homogeneity of the observed specific SFR in high redshift galaxies \citep{elbaz2007a,elbaz2011a,nordon2012a} suggests that the starburst regime might be infrequent, with most galaxies being on the main sequence at any given time. The MIRAGE galaxies always lie on the main sequence, which is consistent and suggests that they do not represent rare objects but rather common galaxies. 

One of the main findings of \cite{perret2014a} is the variation of the SFR on short timescales, with a dispersion $\mathrm{\sigma_{SFR}\simeq0.3\times\left<SFR\right>}$. They attribute this to the presence of multiple long--lived giant star--forming clumps wandering in a turbulent ISM. If they are indeed widespread, these rapid variations of the SFR show the importance of fully understanding what their impact on classical SFR estimators is.

There are two caveats that should nevertheless be noted. First, because these simulations are computationally very demanding, galaxies were followed over less than 1~Gyr. This means that we do not have access to SFH on longer timescales, which prevents us from modelling the contribution of an old stellar population. Cosmological simulations naturally provide SFH over 13~Gyr if they are ran until $z=0$, but by definition such simulations \citep[e.g.][]{pontzen2012a,dubois2014a,vogelsberger2014a} cannot reach the numerical resolution achieved in idealised simulations \citep{renaud2013a}. A numerical resolution that allows to resolve the vertical structure of the disk combined to a gas treatment that allows cooling below $10^4$~K is essential to understand the nature of star formation. Hydrodynamical simulations of a cosmological volume are able to display a cosmic SFH coherent with observations \citep{hopkins2006b}, but at the scale of individual galaxies the SFH is smoothed because of the lack of both spatial and temporal resolution. Using idealised simulations in this analysis, we trade off a realistic long term SFH for more realistic SFR fluctuations on shorter timescales. The second point is that these simulations do not aim at being representative of all main--sequence galaxies at the aforementioned redshifts. In reality, the global population of high--redshift galaxies is likely to exhibit a broader diversity of SFH than can be explored here. However the sample we have adopted is nevertheless very useful to exemplify some of the issues with the measurement of the SFR, even if the exact amplitude of these issues may, in some circumstances, be dependent on the considered SFH.

\subsection{Simulating observations}

For each of the 23 MIRAGE simulations, we compute the attenuation--free spectra every 1~Myr from the SFH, foregoing all spatial information. To do so, we use the \textsc{cigale} SED modelling code (Boquien et al., Burgarella et al. in prep.). We adopt the stellar population models of \cite{bruzual2003a} with a \cite{chabrier2003a} IMF (initial mass function) between 0.1~M$_\odot$ and 100~M$_\odot$ and a metallicity $Z=0.008$, consistent with the subsolar metallicity of galaxies at the considered redshifts \citep[e.g.][]{yuan2013a}. We will discuss the impact of the metallicity on our results in Sect.~\ref{ssec:metal}. The emission from H\,\textsc{ii} regions includes recombination lines and the nebular continuum via free--free, free--bound, and two--photon mechanisms.

The luminosity in individual bands is computed by integrating the spectrum in the corresponding passbands except for the number of Lyman continuum photons which is an output parameter of the model, and for the TIR luminosity, which is computed by integrating over the entire spectrum, based on the assumption that the galaxy is entirely buried, following \cite{kennicutt1998a}.

\begin{figure*}[!htbp]
 \includegraphics[width=\columnwidth]{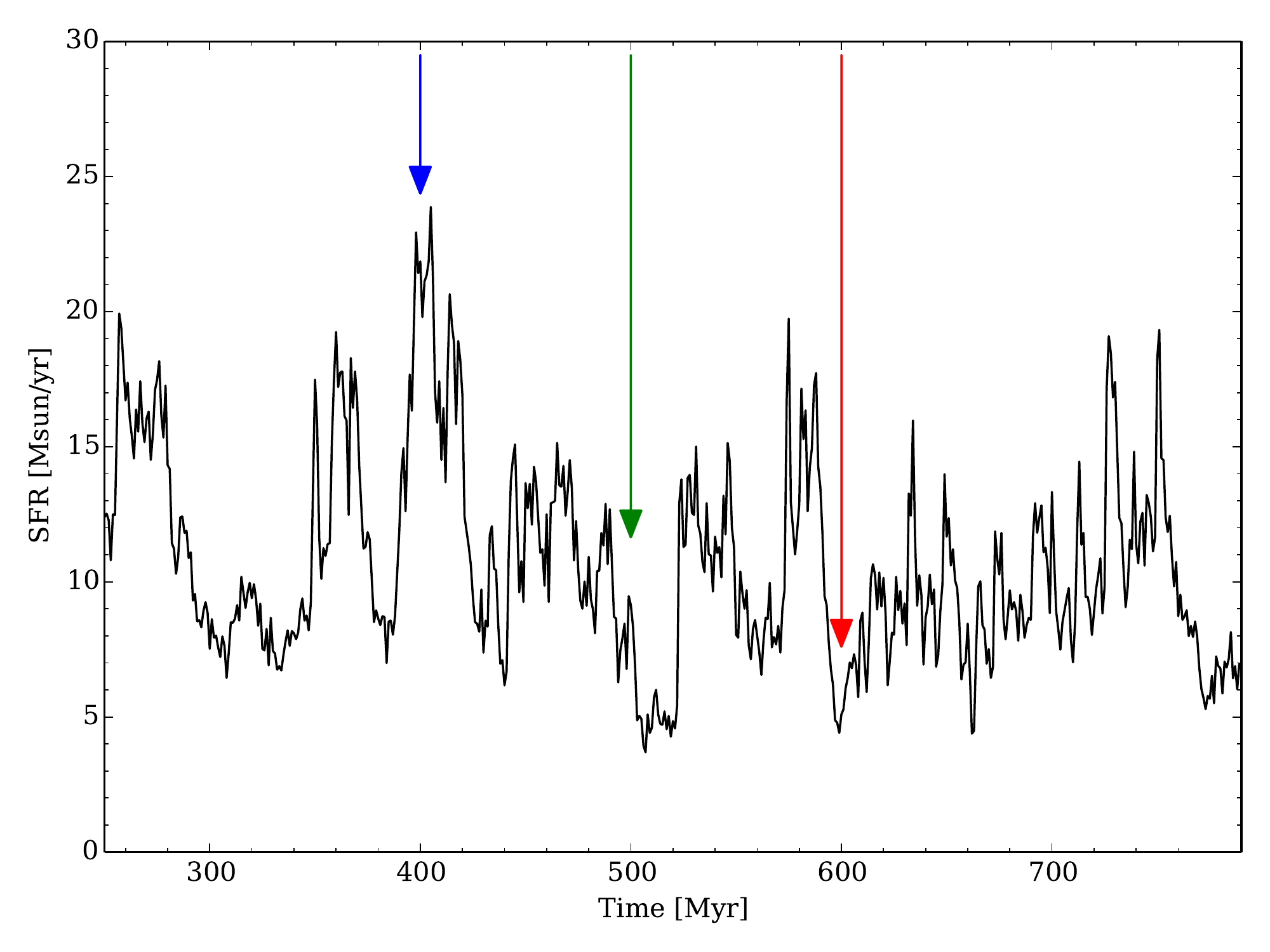}
 \includegraphics[width=\columnwidth]{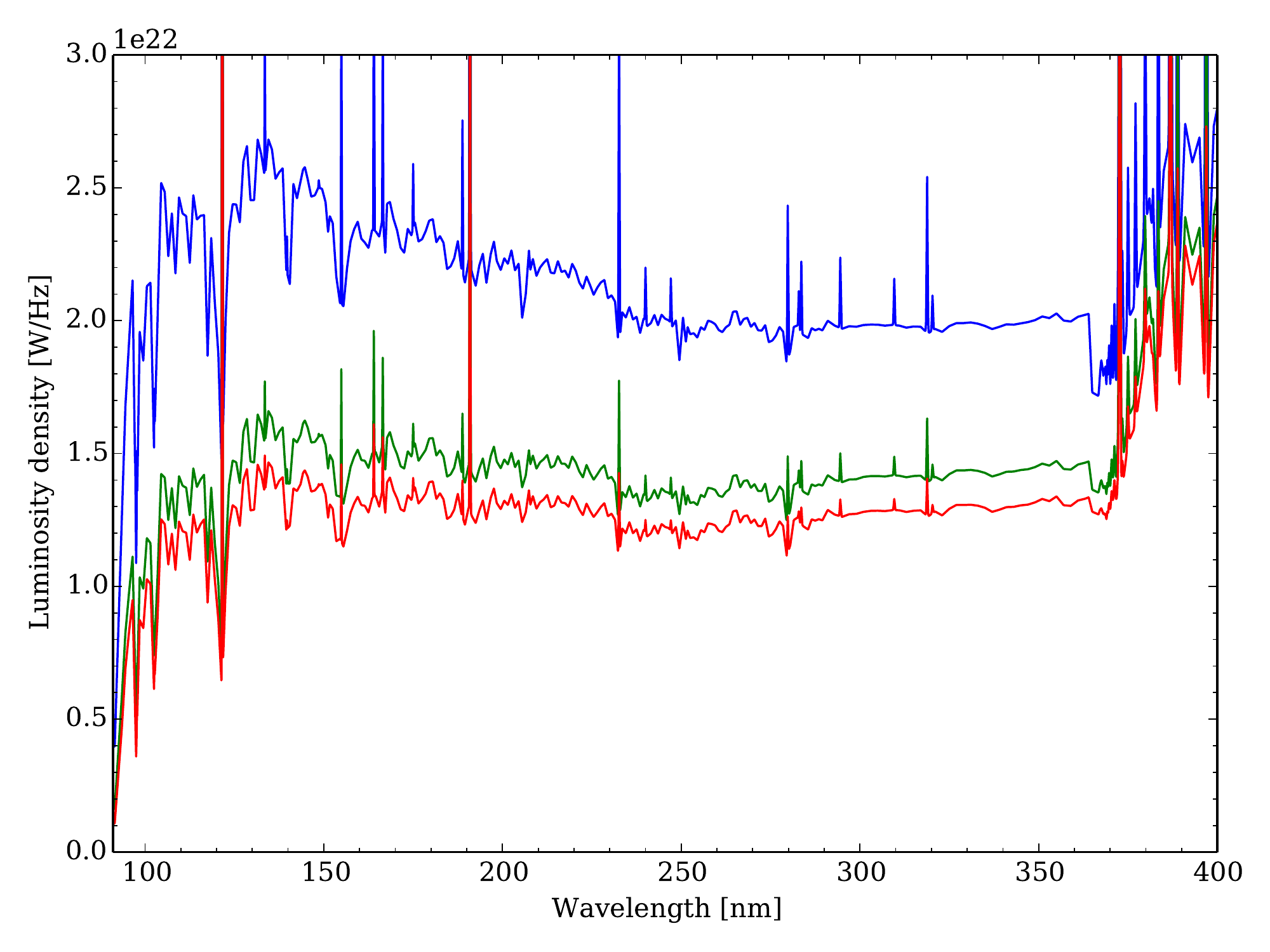}
 \caption{Left: SFH of the MIRAGE G2\_G2\_90\_90\_90 merger simulation \citep{perret2014a}. The time resolution is 1~Myr. The first 250~Myr are not shown as the initial relaxation is handled (Sect.~\ref{ssec:mirage}). The coloured arrows at 400, 500, and 600~Myr indicate at which time we compute the spectra shown in the right--hand panel from the Lyman break to the U band. These spectra are created with the \textsc{cigale} SED modelling code (Boquien et al., Burgarella et al. in prep.). In the present case we assume a \cite{chabrier2003a} IMF and a metallicity $Z=0.008$. Nebular emission is also included. It takes into account recombination lines and the nebular continuum from free--free, free--bound, and two--photon processes. The continuum contributes most in the red part of the shown spectra. We have computed the spectra for all 23 MIRAGE simulations (3 isolated galaxies and 20 interacting galaxies) every 1~Myr.\label{fig:sfh-sed}}
\end{figure*}

An example of a MIRAGE merger SFH and the corresponding spectra from the Lyman break to the U band is shown in Fig.~\ref{fig:sfh-sed} for three different times.
While the details of the SFH differ somewhat from one simulation to another, for instance in terms of normalisation, they are qualitatively similar. There is no long term trend after the initial star formation episode due to the relaxation and peak--to--peak variations are of the order of a few. Overall, the SFH shown in Fig.~\ref{fig:sfh-sed} is typical of the MIRAGE SFH. In terms of relative scatter we find $\mathrm{\sigma_{SFR}\simeq0.36\times\left<SFR\right>}$ versus $0.34\pm0.06$ for the entire sample starting from 250~Myr. All the SFH are shown in Fig.~7 in \cite{perret2014a}.

\subsection{Star formation rate estimators\label{ssec:SFR}}

\subsubsection{Converting a luminosity to an SFR\label{sssec:conv-SFR}}

The computation of the SFR from the luminosity in a given band relies on several assumptions such as a proper correction for partial obscuration by dust, an IMF, a metallicity, and as we have mentioned in Sect.~\ref{sec:intro}, an SFH. Because we want to completely isolate effects due to the variability of the SFH from other effects, the simulated observations and the SFR estimators must be generated with the exact same set of parameters and models. We therefore refrain from adopting SFR estimators from the literature but we rather devise a fully consistent set of estimators. As we will see in Sect.~\ref{sssec:sfr-timescales}, different SFR estimators probe different timescales. A constant SFR over 100~Myr or 1~Gyr is often assumed \citep[e.g.,][]{hao2011a}. We choose the former as our baseline because it typically corresponds to the age at which a constant SFR produces 90\% of the steady state luminosity \citep{kennicutt2012a, boissier2013a} and it is widely used in the literature.

To summarise, we compute the SFR estimators building a synthetic model with the following assumptions: 1) constant SFR over 100~Myr, 2) a \cite{chabrier2003a} IMF between 0.1~M$_\odot$ and 100~M$_\odot$, 3) a metallicity $Z=0.008$. The calibration coefficients are presented in Appendix~\ref{sec:sfr-estimators}. Conceptually similar SFR estimators spanning a broad range in terms of age and metallicity have been presented in \cite{oti2010a}.

In the context of this study, we consider 5 of the most popular SFR estimators: the GALEX FUV, GALEX NUV, and Bessel U bands which trace the photospheric emission from massive stars, the Lyman continuum which is responsible for ionising the hydrogen in star--forming regions and which can be traced through recombination lines or free--free emission, and the total IR emission from dust in a completely buried galaxy.

\subsubsection{A short note on timescales probed by SFR estimators\label{sssec:sfr-timescales}}

\begin{figure}[!htbp]
 \includegraphics[width=\columnwidth]{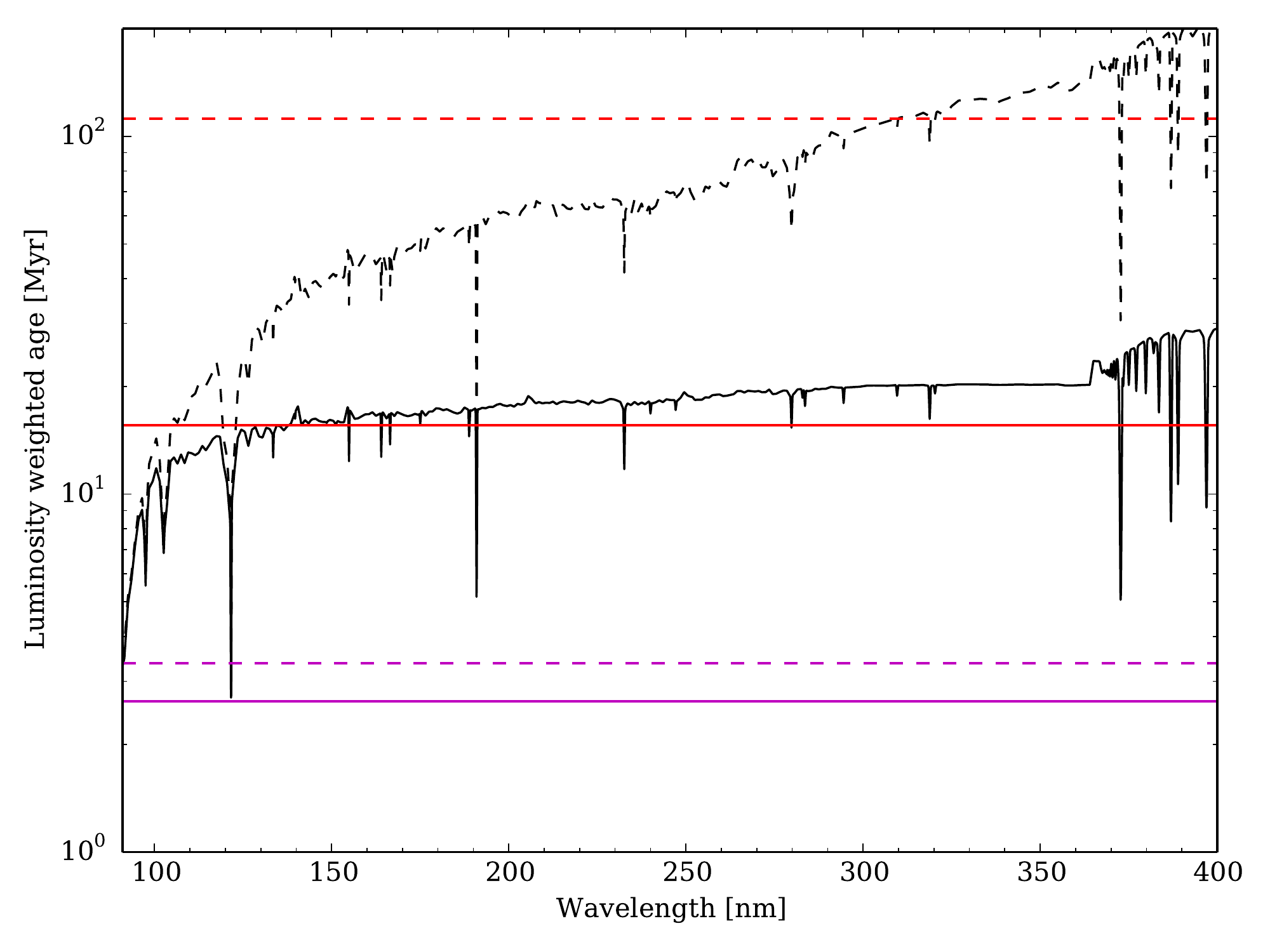}
 \caption{Luminosity weighted ages versus the wavelength for a constant SFR over 100~Myr (solid lines) and 1~Gyr (dashed lines). The magenta (respectively red) line indicates the luminosity weighted age for the Lyman continuum (resp. the TIR for a completely buried galaxy). We see that the typical timescale probed is not only strongly dependent on the estimator and on the wavelength, but also on the SFH, with longer wavelengths and the TIR being the most sensitive to the duration of star formation episode in this simple scenario.\label{fig:t-sfr-cst}}
\end{figure}

Different SFR estimators are generally not sensitive to the same stellar populations. For instance recombination lines are due to massive stars that live of the order of 10~Myr. However, stars that dominate the UV domain live up to a few hundred Myr. Yet, these timescales should rather be seen as an indication of the time required to reach a luminosity steady state for a constant SFR. To quantify what is the typical age of stars dominating the luminosity in a band for a given SFH, we introduce luminosity weighted ages: $\frac{\sum_t t\times L(t)}{\sum_t L(t)}$, with $t$ the age and $L(t)$ the luminosity of a simple stellar population of age $t$ in a given band. In practice we sum over $t$, with each $t$ representing a bin of 1~Myr. In Fig.~\ref{fig:t-sfr-cst}, we present the luminosity weighted ages versus the wavelength for a constant SFR over 100~Myr and over 1~Gyr.
We find that there is a strong dependence of the luminosity weighted ages on the length of the star formation episodes. Unsurprisingly, an episode of 1~Gyr systematically leads to larger ages than in the case of 100~Myr. This also depends on wavelength: longer wavelength bands yield larger ages and on average the TIR is more sensitive to older stars than the UV or the Lyman continuum. This shows the importance of older stellar populations for dust heating when considering timescales longer than 100~Myr, in line with well--known results \citep[e.g.,][]{lonsdale1987a,sauvage1992a,bendo2010a,bendo2012a,boquien2011a,crocker2013a}.

Even for a na\"ive SFH, we see the complex relation there is between 1) the length of the star formation episode, 2) which stars contribute most to a given tracer, and 3) the strong differences from one tracer to another in terms of age sensitivity. More realistic SFH certainly render these relations even more complex. This shows that the presence of long--lived stars also has an impact on SFR estimates. We will examine these issues in detail over the following Sections.

\section{Results\label{sec:results}}

\subsection{Comparison of estimated and true star formation rates\label{ssec:comparison}}

\begin{figure*}[!htbp]
 \includegraphics[width=\columnwidth]{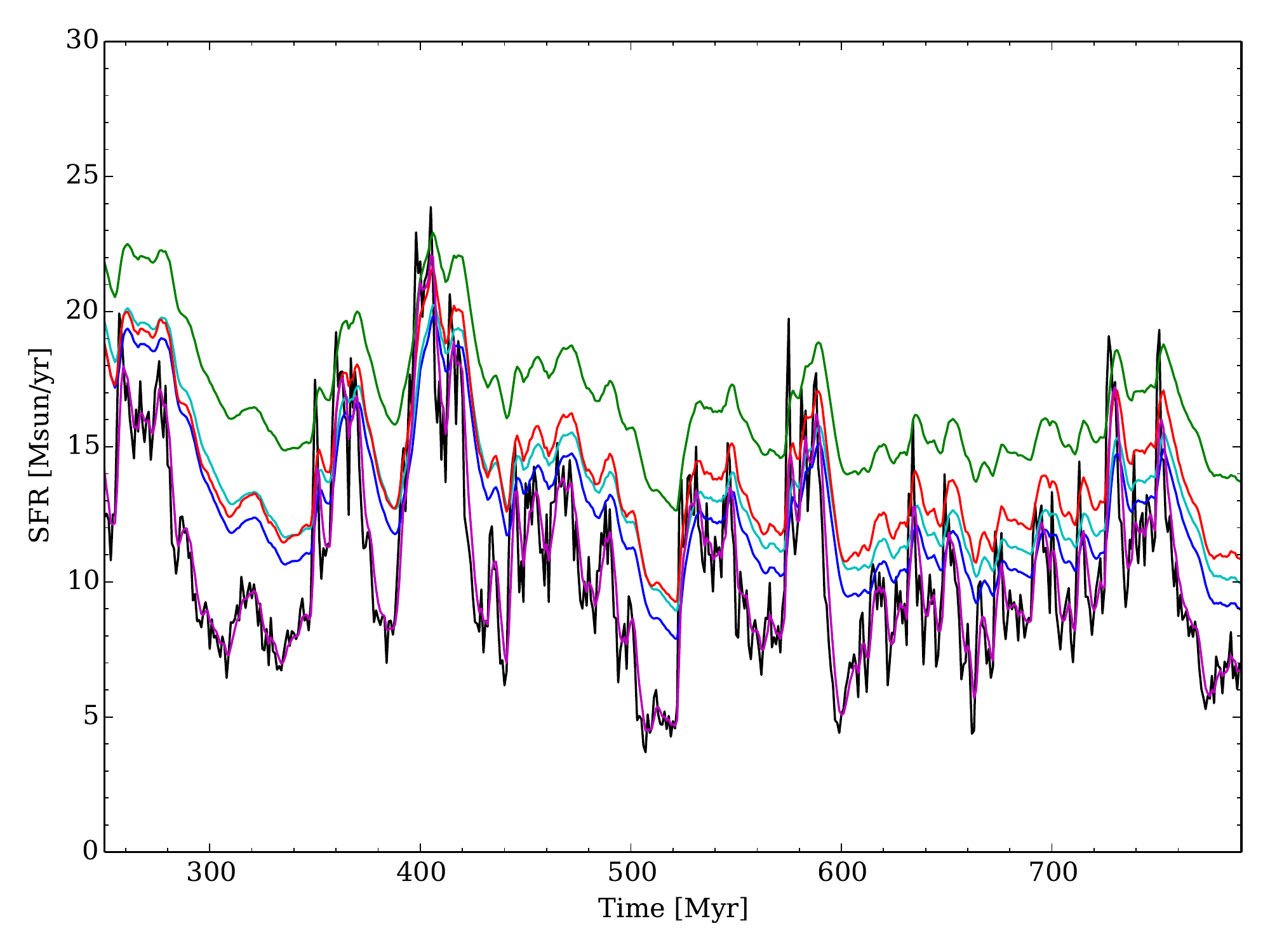}
 \includegraphics[width=\columnwidth]{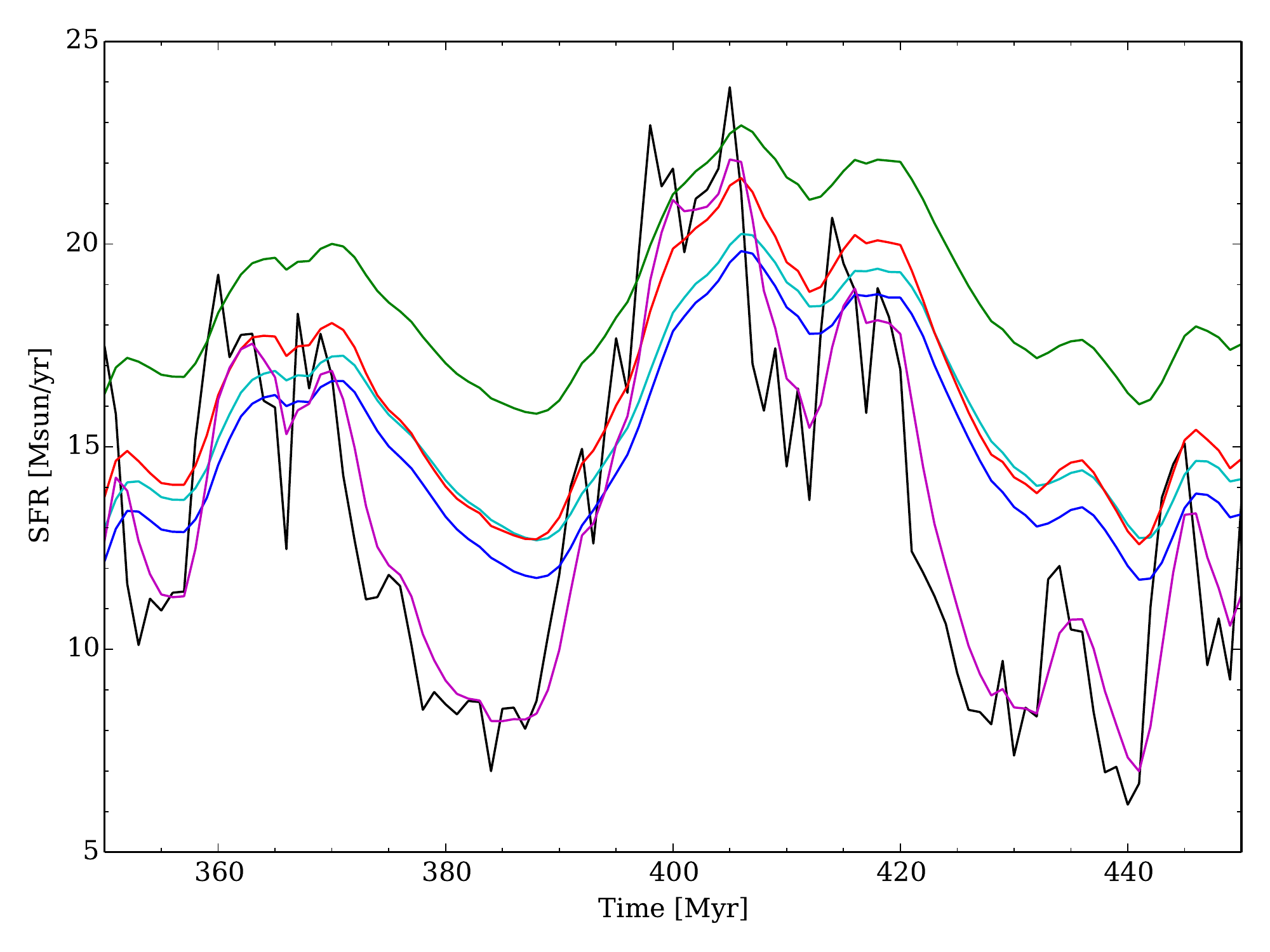}
 \caption{Left: SFR(true) (black), SFR(Ly) (magenta), SFR(FUV) (blue), SFR(NUV) (cyan), SFR(U) (green), and SFR(TIR) (red) for the MIRAGE G2\_G2\_90\_90\_90 simulation \citep{perret2014a}. Note that SFR(true) and SFR(Ly) are nearly blended. Right: zoom between 350~Myr and 450~Myr. We see that SFR(Ly) follows extremely well the variations of SFR(true) even for high frequency variations. SFR estimators at longer wavelengths are not able to capture these variations though, leading to important discrepancies on the estimates of the instantaneous SFR. Beyond the capture of short term variations, we see that there is also a systematic offset that is visible in the FUV and NUV bands but which is especially prominent in the U band. For the present simulation we have $\left<\mathrm{\left[SFR(FUV)-SFR(true)\right]/SFR(true)}\right>=0.27\pm0.30$ between 250~Myr and 790~Myr with a peak of $\sim1.5$. For the U band we have $\left<\mathrm{\left[SFR(U)-SFR(true)\right]/SFR(true)}\right>=0.70\pm0.46$ with a peak of $\sim2.8$. This is representative of the entire MIRAGE sample as can be seen in Fig.~\ref{fig:mean-bias}.\label{fig:sfh-estimated}}
\end{figure*}

\begin{figure}[!htbp]
 \includegraphics[width=\columnwidth]{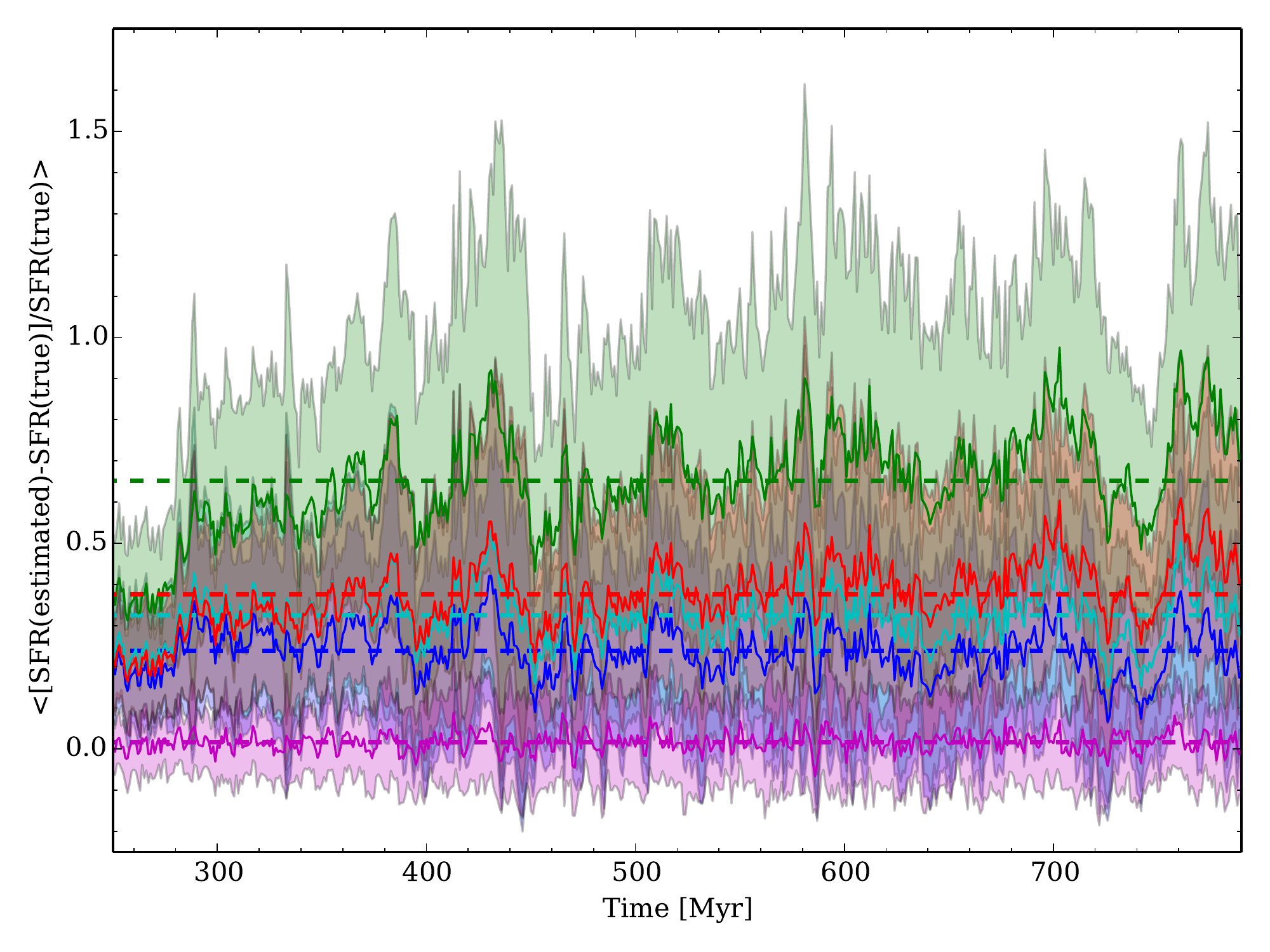}
 \caption{Ratio of difference between SFR(estimated) and SFR(true) to SFR(true) versus time for the entire MIRAGE sample. The solid lines indicate the mean value across all simulations whereas the shaded areas of the same colour indicate the standard deviation. The dashed lines indicate the mean value between 250~Myr and 790~Myr. The colour coding is the same as for Fig.~\ref{fig:sfh-estimated}. We see that except for the Lyman continuum, the SFR is systematically overestimated, from 24\% on average for the FUV band to 65\% on average for the U band. Peak relative differences reach $\sim0.7$ for the Lyman continuum, $\sim1.9$, $\sim2.1$, and $\sim3.2$ for the FUV, NUV, and U bands, and $\sim2.0$ for the TIR.\label{fig:mean-bias}}
\end{figure}

For each of the 23 simulations, we have estimated the SFR using the conversion factors devised in Sect.~\ref{sssec:conv-SFR} and listed in Appendix~\ref{sec:sfr-estimators} which assume a metallicity $Z=0.008$, a \cite{chabrier2003a} IMF between 0.1~M$_\odot$ and 100~M$_\odot$, and a constant SFR over 100~Myr. We compare SFR(estimated) with SFR(true) for one of the simulations in Fig.~\ref{fig:sfh-estimated}.
We find that SFR(Ly) follows very closely SFR(true) and is able to capture all but the highest frequency variations for which it suffers from a short time delay. Conversely, SFR(FUV), SFR(NUV), SFR(U), and SFR(TIR) do not properly capture these variations lasting no more than a few tens Myr. A related and important question is whether the SFR(estimated) are accurate on average on long timescales for which the short term variations could be averaged out. We find that in reality SFR(estimated) systematically overestimate SFR(true). If the overestimate is very small for SFR(Ly) it is clearly visible for SFR(FUV) and SFR(NUV), and it is particularly strong for SFR(U): we find $\left<\mathrm{\left[SFR(FUV)-SFR(true)\right]/SFR(true)}\right>=0.27\pm0.30$ and  $\left<\mathrm{\left[SFR(U)-SFR(true)\right]/SFR(true)}\right>=0.70\pm0.46$ between 250~Myr and 790~Myr (we forego the first 250~Myr to avoid being affected by the highly transitory nature of the onset of star formation). To examine this issue further, in Fig.~\ref{fig:mean-bias} we compare the ratio of the difference between SFR(estimated) and SFR(true) to SFR(true) versus time for the entire sample. 
We find that this excess can also be retrieved when considering the entire sample:
\begin{itemize}
 \item $\left<\mathrm{\left[SFR(Ly) -SFR(true)\right]/SFR(true)}\right> = 0.02\pm0.11$
 \item $\left<\mathrm{\left[SFR(FUV)-SFR(true)\right]/SFR(true)}\right> = 0.24\pm0.26$
 \item $\left<\mathrm{\left[SFR(NUV)-SFR(true)\right]/SFR(true)}\right> = 0.33\pm0.29$
 \item $\left<\mathrm{\left[SFR(U)  -SFR(true)\right]/SFR(true)}\right> = 0.65\pm0.42$
 \item $\left<\mathrm{\left[SFR(TIR)-SFR(true)\right]/SFR(true)}\right> = 0.38\pm0.29$
\end{itemize}
This means that except for the Lyman continuum, classical estimators calibrated assuming a constant SFR over 100~Myr all clearly overestimate the SFR. We will discuss the origin of this effect in Sect.~\ref{ssec:origin} and its impact in Sect.~\ref{ssec:impact}.

\subsection{Timescales}

\begin{figure*}[!htbp]
 \includegraphics[width=\columnwidth]{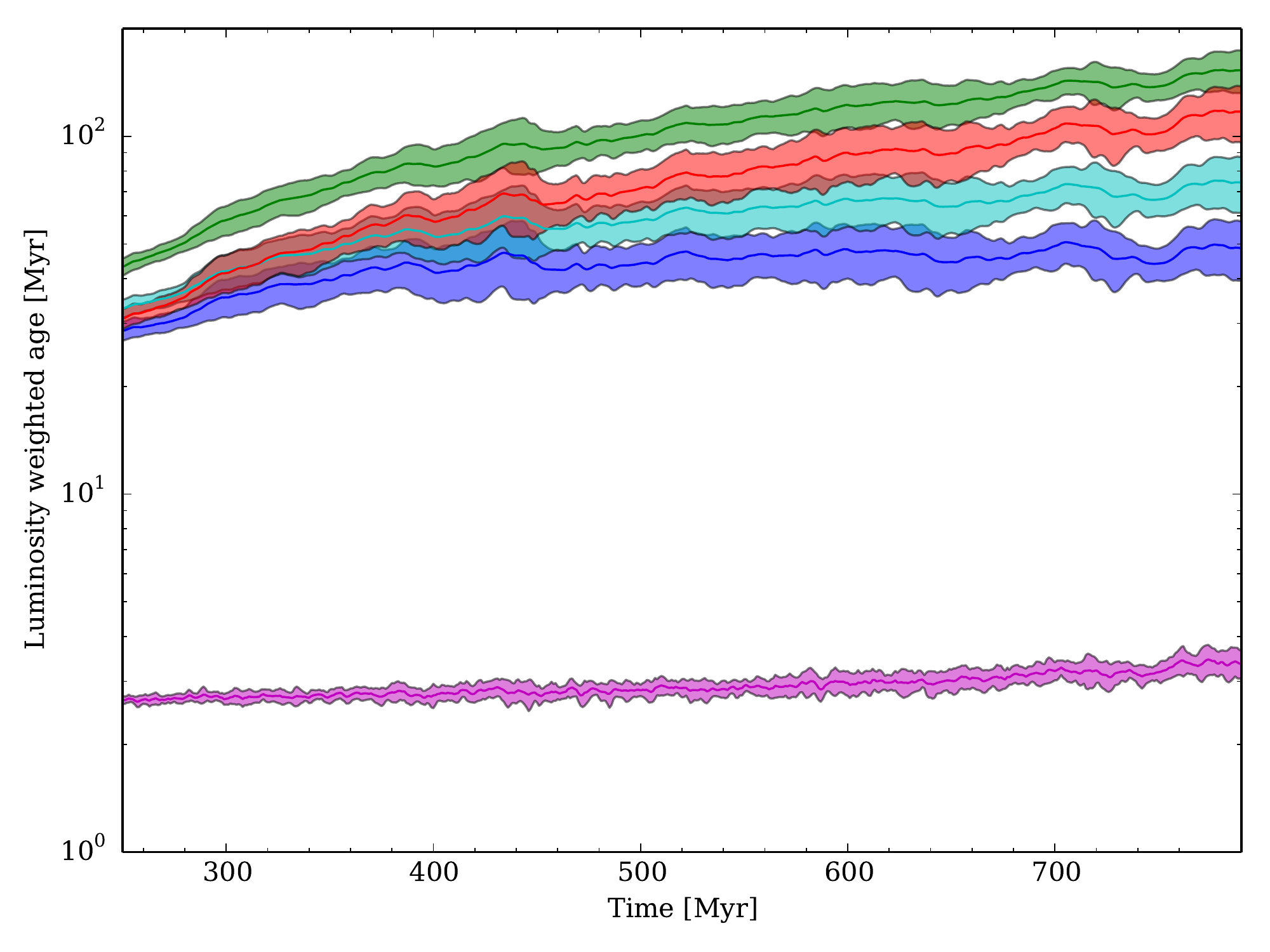}
 \includegraphics[width=\columnwidth]{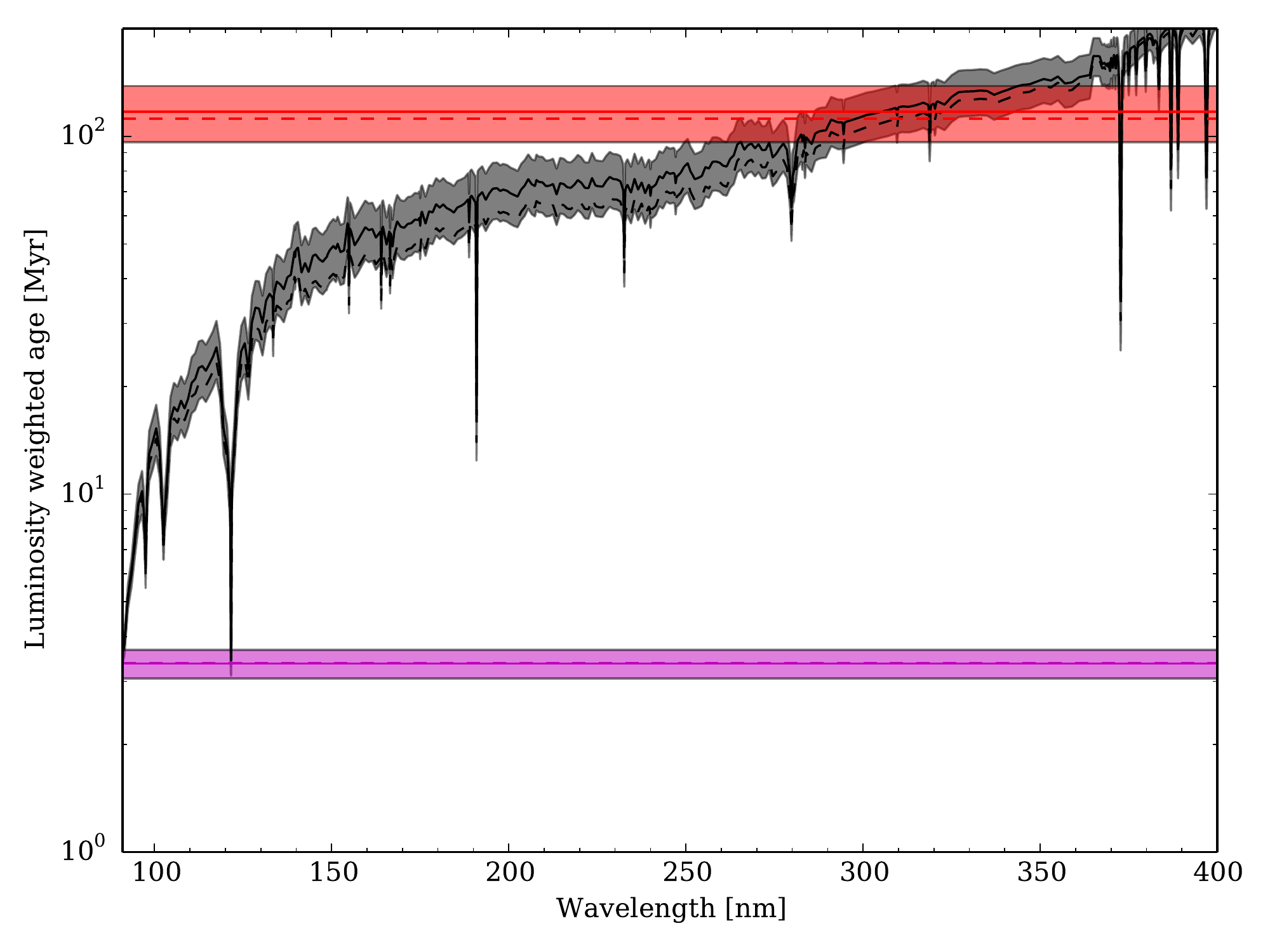}
 \caption{Left: Luminosity weighted ages versus the simulation time. The solid lines indicate the mean value of for the 23 simulations and the shaded areas of the corresponding colours indicate the standard deviation. The colour scheme is the same as that of Fig.~\ref{fig:sfh-estimated}. Right: Luminosity weighted ages versus wavelength (black) for a simulation age of 790~Myr. The values for the Lyman continuum and the TIR at 790~Myr are also indicated. For comparison, the corresponding values for a constant SFR over 1~Gyr shown in Fig.~\ref{fig:t-sfr-cst} is indicated in dashed lines.\label{fig:timescales-sfh}}
\end{figure*}

We have seen in Sect.~\ref{sssec:sfr-timescales} that the typical timescales probed by different estimators are strongly dependent on the duration of the star formation episode. However, we examined the case of a constant SFR over a long period. It is unclear how these timescales compare with those for more realistic SFH. In Fig.~\ref{fig:timescales-sfh}, we have computed the luminosity weighted ages versus the simulation age and versus the wavelength.
There is a generally progressive increase of the luminosity weighted ages with time. However, after 400~Myr, FUV progressively reaches a steady state. When considering longer wavelengths, the weighted ages keep increasing because of the continuing accumulation of long--lived stars. Short term variations of the SFR seem to have little overall impact. At 790~Myr, the Lyman continuum has a luminosity weighted age of $\sim3$~Myr, the FUV band $\sim40-60$~Myr, the NUV band $\sim60-90$~Myr, and the TIR and the U band tracers more than 100~Myr. Finally, when comparing to Fig.~\ref{fig:t-sfr-cst}, we find that the dependence of the ages on the wavelength is remarkably similar to the case of a constant SFR over 1~Gyr.

\section{Discussion\label{sec:discussion}}

\subsection{Why the SFR is overestimated\label{ssec:origin}}

We have found that except for the Lyman continuum, other SFR estimators systematically overestimate SFR(true) over long periods, from 24\% on average for the FUV band up to 65\% for the U band. We examine here the fundamental reasons for this discrepancy and under which conditions standard estimators could provide reasonably unbiased results.

\begin{figure*}[!htbp]
 \includegraphics[width=\columnwidth]{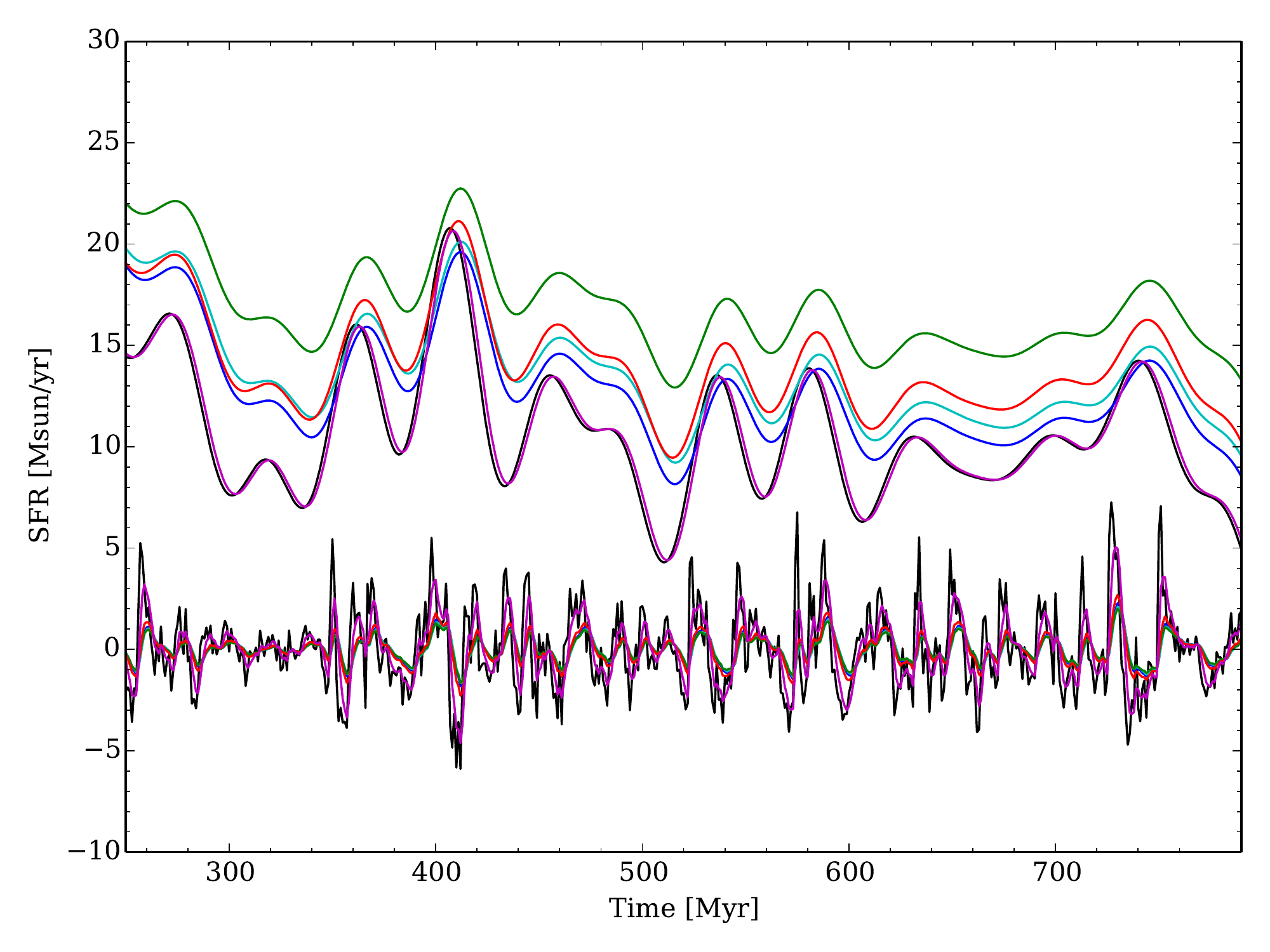}
 \includegraphics[width=\columnwidth]{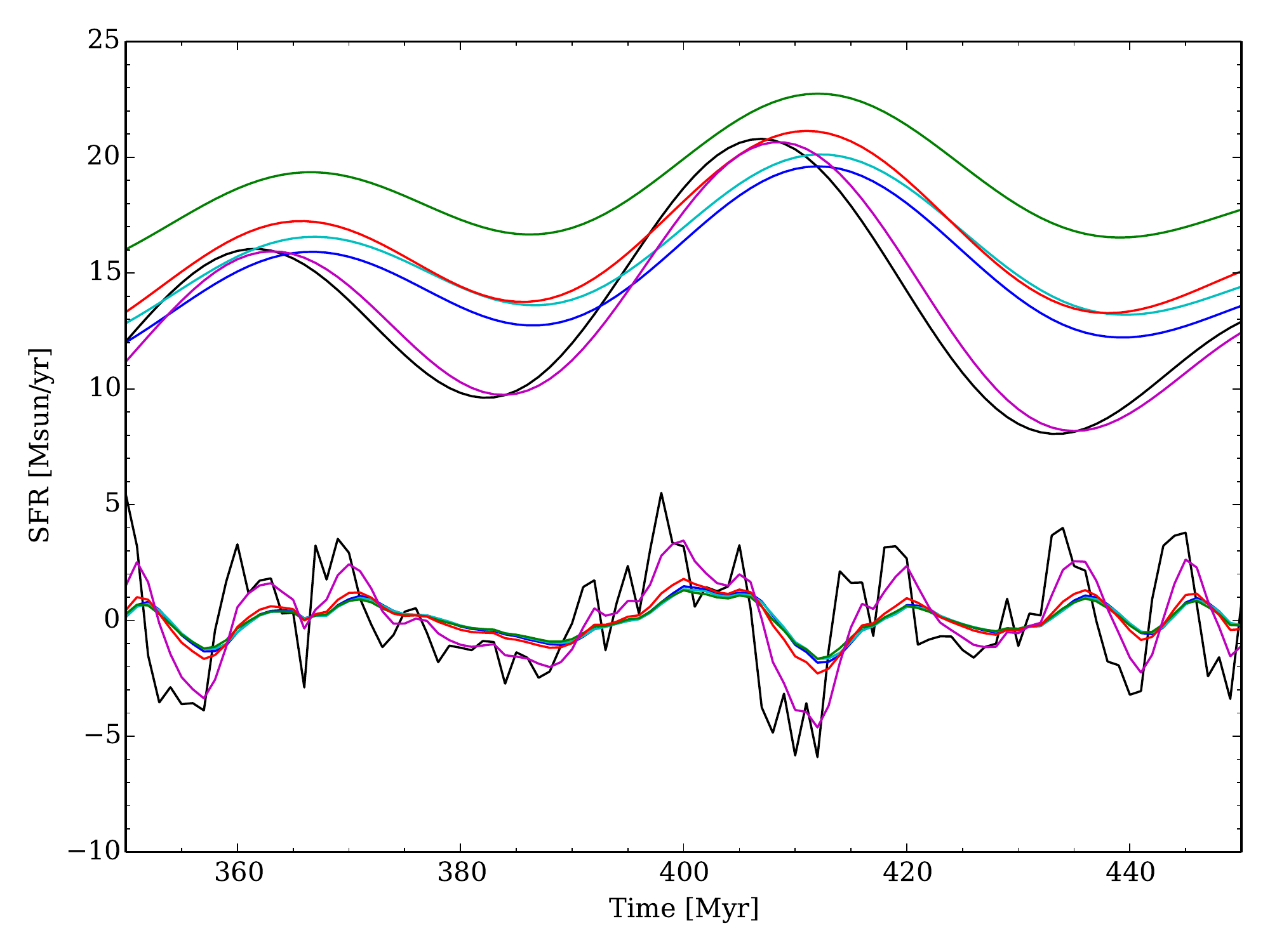}
 \caption{Left: Fourier--filtered SFH of the MIRAGE G2\_G2\_90\_90\_90 simulation (black lines) along with SFR(estimated) recomputed from each of the two filtered SFH.The top (resp. bottom) lines represent the low (resp. high) frequency component. The colours follow the same scheme as that of Fig.~\ref{fig:sfh-estimated}. Note that SFR(true) and SFR(Ly) are nearly blended. Right: zoom between 350~Myr and 450~Myr. We find that if there can be important discrepancies in the SFR, on average at high frequency SFR(true) and SFR(estimated) give similar results, both close to 0~Msun/yr. The behaviour at low frequency is markedly different. We see that SFR(estimated) show both a systematic overestimate and a delay in time from the extrema of SFR(true). The smallest delay occurs for SFR(Ly). Conversely, SFR(FUV), SFR(NUV), SFR(U), and SFR(TIR) all show approximately similar delays of a few Myr.\label{fig:sed-fourier}}
\end{figure*}

Intuitively, SFR estimators sensitive to short timescales such as SFR(Ly) provide us with accurate estimates because they tightly follow the variations of SFR(true) both when it increases and when it decreases, albeit with a slight delay. However, as other tracers are sensitive to longer timescales we are facing the combined impacts of short and long term variations of the SFR. To investigate these issues, we have decomposed the MIRAGE SFH into low and high frequency components using a Fast Fourier Transform. While the choice of the threshold in terms of variation period between these components is somewhat arbitrary, we have chosen a separation period of 40 Myr, as it is similar to the FUV age--weighted luminosities. That is to say that all variations faster (resp. slower) than a timescale of 40~Myr are assigned to the high (resp. low) frequency component. From each of these two SFH we have recomputed SFR(estimated) independently. We show the Fourier--filtered SFH along with the corresponding recomputed SFR(estimated) for one of the MIRAGE simulations in Fig.~\ref{fig:sed-fourier}. Note that the SFR(estimated) are not Fourier--filtered, they are only computed from Fourier--filtered SFH.

First, we only consider the high frequency component. We see that only SFR(Ly) can follow the rapid variations of SFR(true). But even then, there is always a slight delay. Other tracers cannot follow rapid variations accurately but no systematic bias is visible. If we look at the entire sample between 250~Myr and 790~Myr, we find:
\begin{itemize}
 \item $\left<\mathrm{\left[SFR_{HF}(Ly)-SFR_{HF}(true) \right]/SFR(true)}\right>=0.01\pm0.11$
 \item $\left<\mathrm{\left[SFR_{HF}(FUV)-SFR_{HF}(true)\right]/SFR(true)}\right>=0.03\pm0.17$
 \item $\left<\mathrm{\left[SFR_{HF}(NUV)-SFR_{HF}(true)\right]/SFR(true)}\right>=0.03\pm0.18$
 \item $\left<\mathrm{\left[SFR_{HF}(U)-SFR_{HF}(true)  \right]/SFR(true)}\right>=0.03\pm0.17$
 \item $\left<\mathrm{\left[SFR_{HF}(TIR)-SFR_{HF}(true)\right]/SFR(true)}\right>=0.03\pm0.16$
\end{itemize}
On average, the bias due to rapid variations of the SFR is limited to 3\% and even just 1\% for SFR(Ly). However, rapid variations of the SFR increase the scatter on the instantaneous SFR as even the SFR(Ly) cannot follow the variations closely enough. If we compare these values with those obtained in Sect.~\ref{ssec:comparison}, it appears that rapid variations are responsible for nearly all of the scatter for SFR(Ly). However for other tracers, this is not true as the scatter due to rapid variations of the SFR alone cannot explain all of the total scatter. This suggests that low frequency variations play an important role.

\begin{figure}[!htbp]
 \includegraphics[width=\columnwidth]{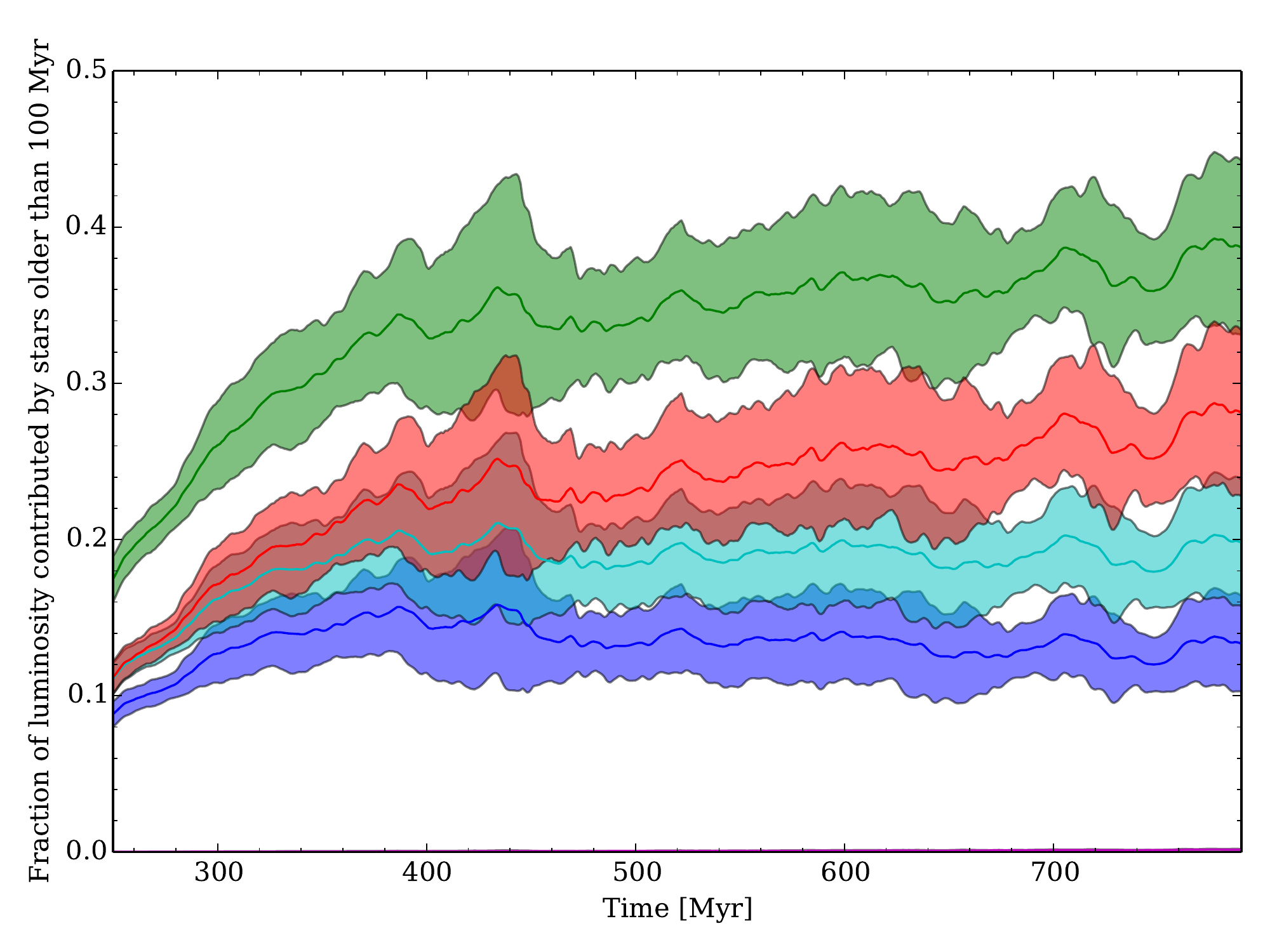}
 \caption{Fraction of the luminosity in star formation tracing bands contributed by stars living over 100~Myr versus simulation time. The colour scheme is the same as that of Fig.~\ref{fig:sfh-estimated}. The contribution of long--lived stars is negligible for the Lyman continuum but reaches upward 10\% in the FUV, a small but non--negligible fraction. The contamination is stronger for the NUV and TIR tracers, and it is maximum for the U band with a typical value of $\sim35$\%.\label{fig:contrib_old}}
\end{figure}

\begin{figure*}[!htbp]
 \includegraphics[width=\columnwidth]{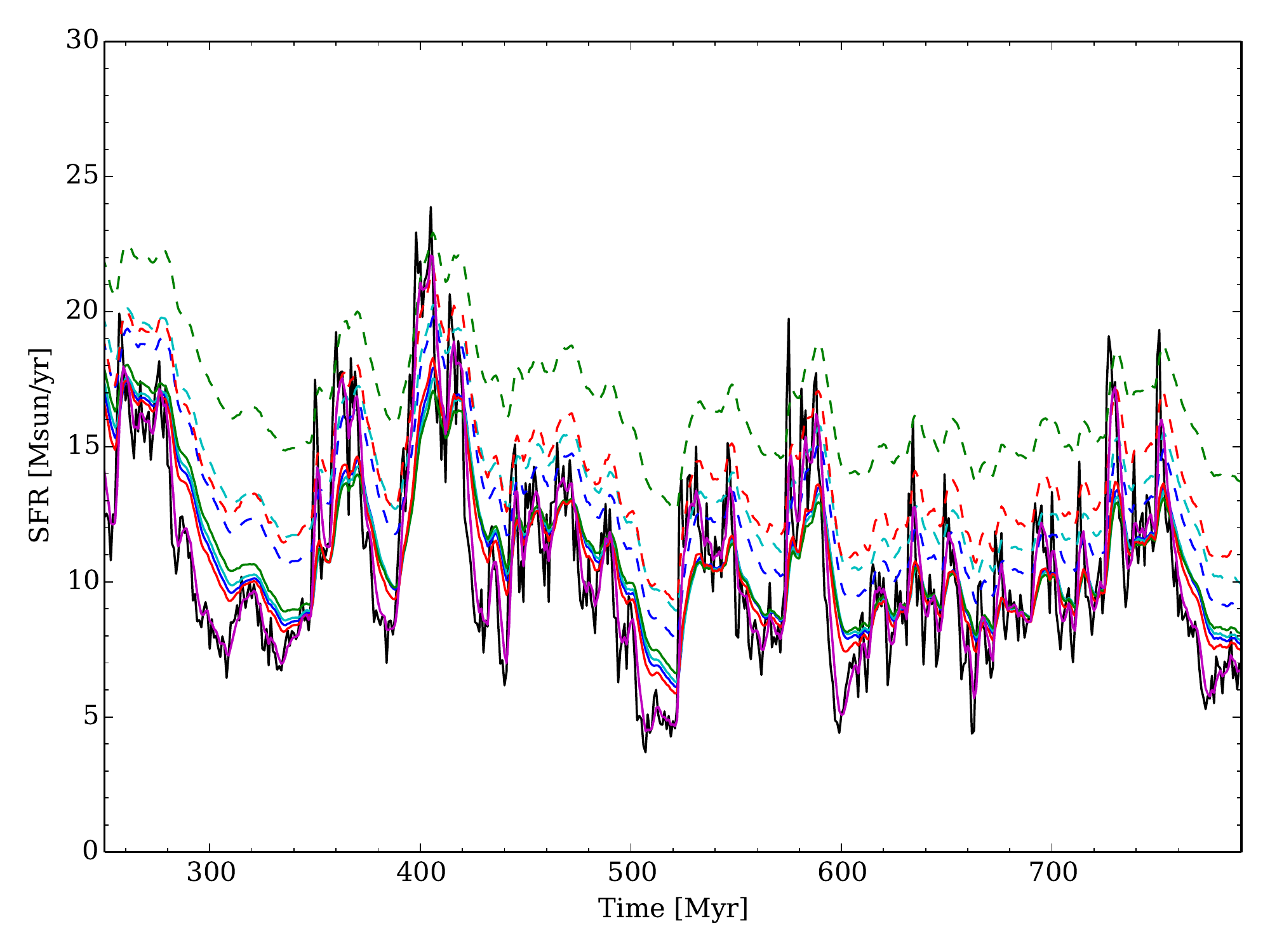}
 \includegraphics[width=\columnwidth]{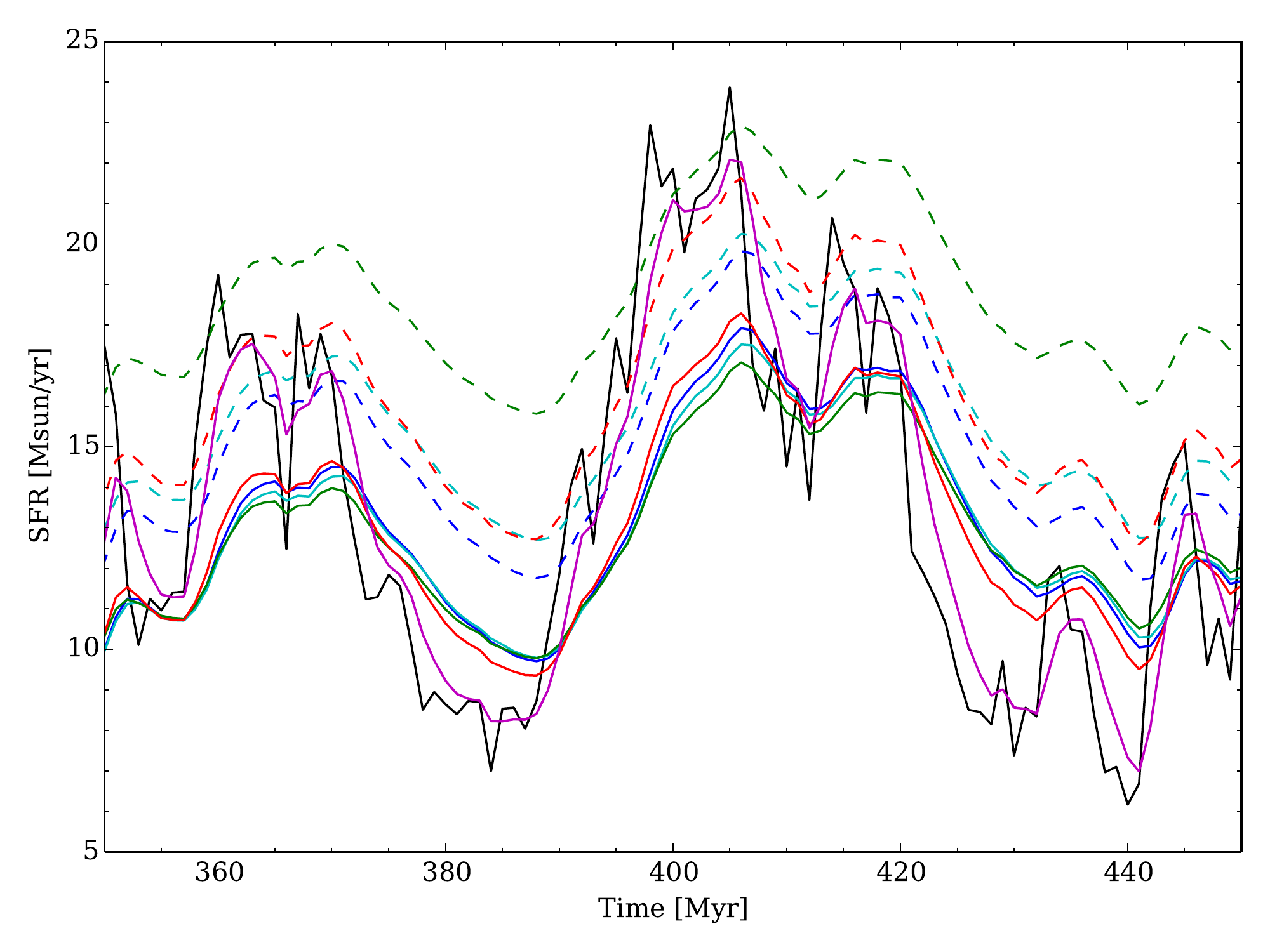}
 \caption{Same as Fig.~\ref{fig:sfh-estimated} with the SFR(estimated) corrected for the contamination by long--lived stars. For reference uncorrected luminosities are indicated with dashed lines. We see that once corrected, SFR(estimated) are in much better agreement with SFR(true) and that a large part of the bias has been eliminated. This is especially visible for the U band.\label{fig:sfh-estimated-noold}}
\end{figure*}

\begin{figure}[!htbp]
 \includegraphics[width=\columnwidth]{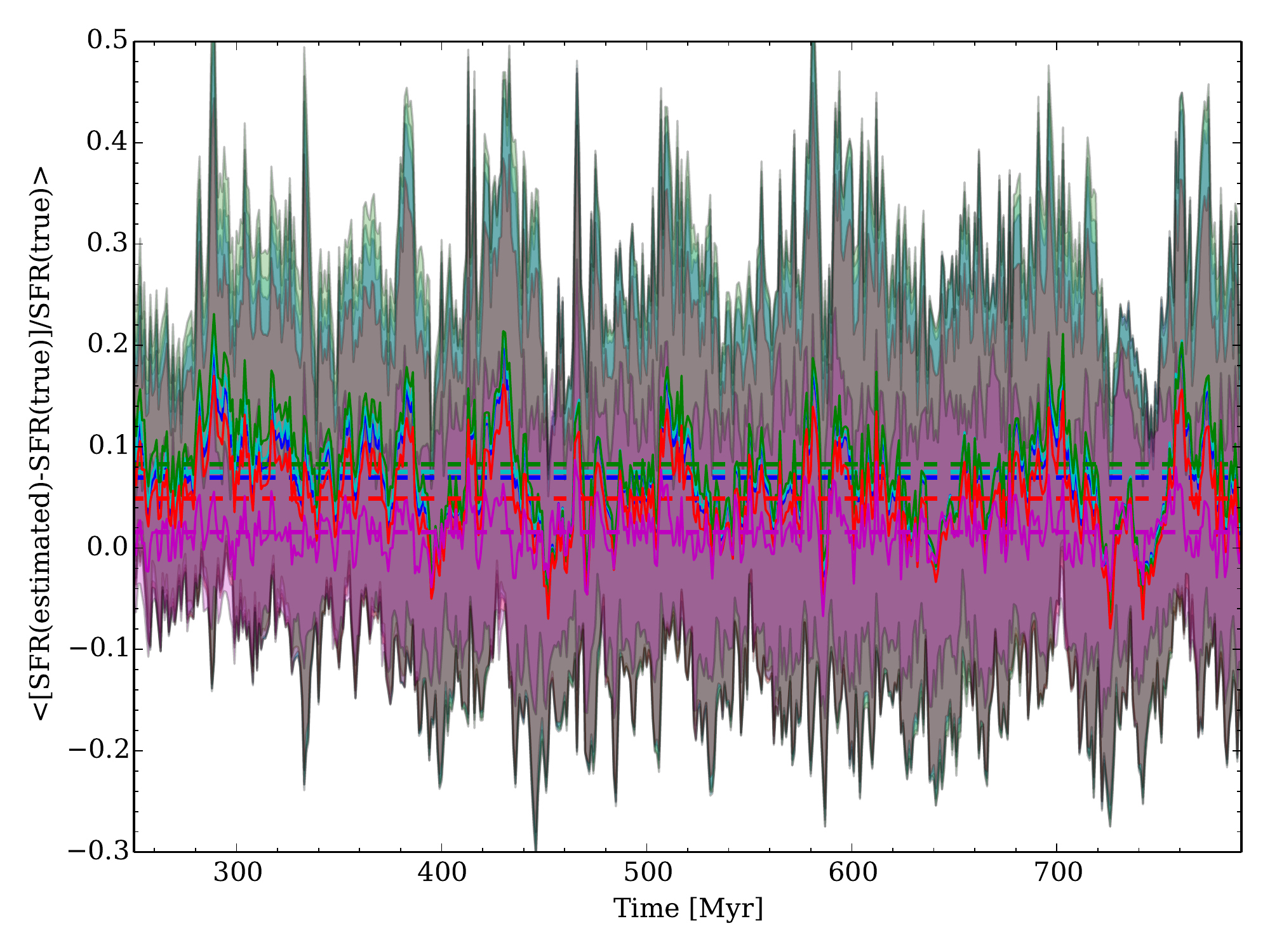}
 \caption{Same as Fig.~\ref{fig:mean-bias} with SFR(estimated) corrected for the contamination by long--lived stars. We see that the bias and the scatter are strongly reduced in all bands. This is especially visible in the case of the U band.\label{fig:mean-bias-noold}}
\end{figure}

If we now consider only the low frequency component, we can see two main features. First, in the right panel of Fig.~\ref{fig:sed-fourier} we see is a clear offset between the extrema of SFR(true) and that of SFR(estimated). This delay is as expected minimal for SFR(Ly), around 1~Myr. For other estimators it is typically delayed by a few Myr. This delay appears remarkably independent of the wavelength from SFR(FUV) to SFR(TIR). The second obvious feature is that except for SFR(Ly) there is an excess in SFR(estimated). It appears that tracers having longer timescales have a stronger excess. If we consider the entire sample, we find:
\begin{itemize}
 \item $\left<\mathrm{\left[SFR_{LF}(Ly)-SFR_{LF}(true)\right)/SFR(true)}\right>=0.00\pm0.03$
 \item $\left<\mathrm{\left[SFR_{LF}(FUV)-SFR_{LF}(true)\right)/SFR(true)}\right>=0.21\pm0.17$
 \item $\left<\mathrm{\left[SFR_{LF}(NUV)-SFR_{LF}(true)\right)/SFR(true)}\right>=0.30\pm0.20$
 \item $\left<\mathrm{\left[SFR_{LF}(U)-SFR_{LF}(true)\right)/SFR(true)}\right>=0.62\pm0.33$
 \item $\left<\mathrm{\left[SFR_{LF}(TIR)-SFR_{LF}(true)\right)/SFR(true)}\right>=0.35\pm0.20$
\end{itemize}
This suggests there is an impact due to the long term variations of the SFR and/or a contamination from stars living longer than the calibration timescale, here 100~Myr. To investigate the latter possibility, we have computed the fraction of the luminosity contributed by such long--lived stars for the entire sample in Fig~\ref{fig:contrib_old}.
We find that except for the Lyman continuum, long--lived stars can account for a sizeable fraction of the luminosity in star formation tracing bands. The typical contribution ranges from slightly more than $\sim10$\% for the FUV band to $\sim35$\% for the U band. While such a contribution is not sufficient to explain the full extent of the bias shown in Sect.~\ref{ssec:comparison}, it already explains at least half of the observed bias. We show SFR(estimated) after correction for the presence of long--lived stars for one MIRAGE simulation in Fig.~\ref{fig:sfh-estimated-noold}.
We find that the SFR are in better agreement once they are corrected for the contamination by stars living longer than 100~Myr. The change is the most spectacular for the U band. To examine this issue further, in Fig.~\ref{fig:mean-bias-noold} we compare the ratio of the difference of contamination--corrected SFR(estimated) with SFR(true) to SFR(true) for the entire sample.
We find that the bias is strongly reduced:
\begin{itemize}
 \item $\left<\mathrm{\left[SFR(Ly)-SFR(true) \right]/SFR(true)}\right> = 0.02\pm0.11$
 \item $\left<\mathrm{\left[SFR(FUV)-SFR(true)\right]/SFR(true)}\right> = 0.07\pm0.20$
 \item $\left<\mathrm{\left[SFR(NUV)-SFR(true)\right]/SFR(true)}\right> = 0.08\pm0.21$
 \item $\left<\mathrm{\left[SFR(U)-SFR(true)  \right]/SFR(true)}\right> = 0.09\pm0.22$
 \item $\left<\mathrm{\left[SFR(TIR)-SFR(true)\right]/SFR(true)}\right> = 0.05\pm0.18$
\end{itemize}
Of the remaining bias, a third to a half of it is due to rapid variations of the SFR. The rest of it is probably due to slower deviations from the assumption of a constant SFR.

As a summary, except for the Lyman continuum, standard SFR estimators calibrated over 100~Myr provide us with biased results: from $\sim25$\% for SFR(FUV) to $\sim65$\% for SFR(U). This is mainly due to the accumulation of stars living over 100~Myr. After correction for long--lived stars, the bias is smaller than 10\% on average. The scatter is due to the combination to short term and long term variations of the SFR. These results are in line with the recent findings of \cite{johnson2013a}. Reconstructing SFH from colour--magnitude diagrammes of a sample of nearby dwarf galaxies, they found that long--lived stars contribute from less than 5\% to 100\% for the FUV luminosity. 

\subsection{Effect of the metallicity\label{ssec:metal}}

We have carried out the present study for $Z=0.008$. However, the metallicity has a direct effect on stellar evolutionary tracks and on stellar atmospheres. To understand the exact impact of the metallicity on our results, we have recomputed SFR(estimated) at different metallicities. All estimators have been systematically recalibrated for each metallicity to ensure the study remains self--consistent. We find that on average, as the metallicity decreases the bias increases. For the FUV (resp. U) band, the relative bias ranges from $0.19\pm0.23$ (resp. $0.57\pm0.38$) for $Z=0.02$ to $0.37\pm0.32$ (resp. $0.87\pm0.52$) for $Z=0.0001$. Because low metallicity stellar populations tend to be bluer than more metal--rich populations of the same age, at low metallicity there is probably an increased accumulation of long--lived stars contributing to star--formation tracing bands. However, for such extreme metallicities, the time from the onset of star formation is necessarily much smaller than at $1<z<2$ so the time available to accumulate long--lived stars is shorter and the metallicity rapidly increases, limiting the impact of the contribution of such stars.

\subsection{Impact on practical measurements of the SFR\label{ssec:impact}}

In the ideal case we have considered, we find that tracers based on the Lyman continuum, such as free--free emission or recombination lines estimate the SFR very accurately, within 2\%. The reason is that tracers based on the Lyman continuum provide the instantaneous SFR, or the rate at which massive stars are being formed within a time lapse of a few Myr, while other tracers are giving the SFR averaged in a complex way over the typical timescale of the stars producing them. If all estimators are affected in reality at various degrees from dust obscuration or assumptions on the IMF, the Lyman continuum faces specific problems. The first issue is that it is difficult to measure observationally in the context of large surveys of high--redshift galaxies, be it from free--free emission for lack of sensitivity or from recombination lines. Indeed, measuring recombination lines requires spectra or specific narrow--band filters which severely limit the redshift range. In addition, at high redshift the Ly$\alpha$ line is the only widely usable line but it is very sensitive to resonant scattering by neutral hydrogen \citep[e.g., ][]{hayes2013a}, inducing large variations in the line strength even at fixed SFR. Another issue that has surfaced in recent years is the sensitivity of SFR calibration to models. In particular, the inclusion of stellar rotation can have a strong impact on the production of Lyman continuum photons \citep{levesque2012a}. If unconstrained such issues may outweigh the strong advantage of the Lyman continuum outlined in this paper, in which we have avoided the issue of model uncertainties on purpose.

\begin{figure}[!htbp]
 \includegraphics[width=\columnwidth]{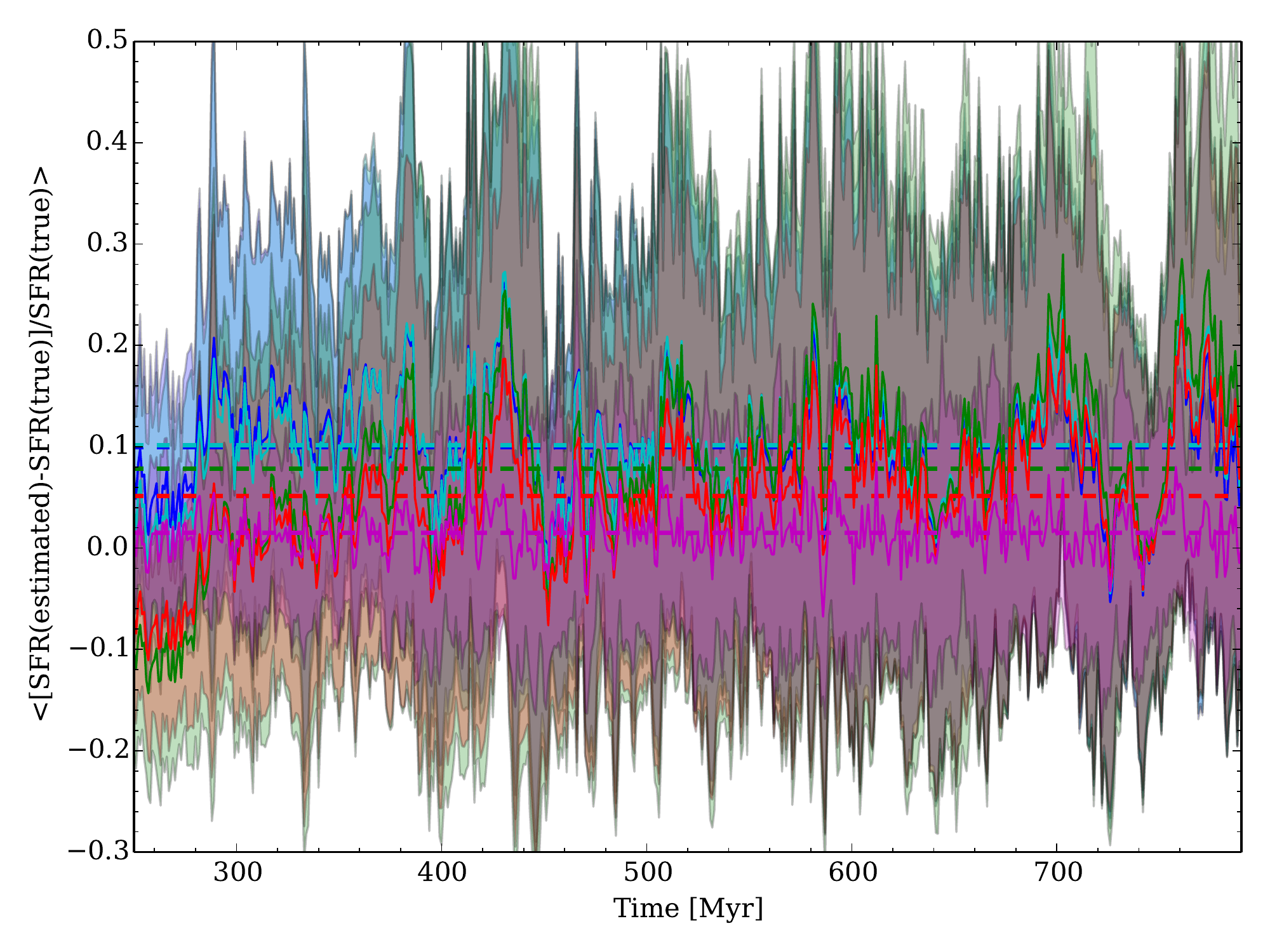}
 \caption{Same as Fig.~\ref{fig:mean-bias} with SFR estimators calibrated over 1~Gyr rather than 100~Myr. We see that this strongly improves the agreement between the SFR(estimated) and SFR(true) and provides estimates nearly as good as when correcting for the presence of stars older than 100~Myr. The remainder of the differences is likely due to variations of the SFR.\label{fig:mean-bias-1000}}
\end{figure}

Even though it is much more sensitive to the variation of the SFR than the Lyman continuum, the FUV emission provides us with the second most accurate SFR estimator of the set we consider. This being said, on average for the simulated SFH it overestimates the SFR by $\sim25$\% on average. The origin of this excess provides us with two important lessons to keep in mind for measuring the SFR in large surveys. First, unless the galaxy is currently undergoing a strong episode of star formation that completely outshines older stellar populations in star formation tracing bands, these long--lived stars need to be taken into account. In the case of the MIRAGE simulations, correcting for the presence of stars living more than 100~Myr lowers the bias by a factor $\sim3$ in the FUV band and by a factor $\sim7$ in the U band. We therefore recommend to adopt SFR estimators calibrated over timescales longer than 100~Myr as is usually done. As we can see in Fig.~\ref{fig:mean-bias-1000}, taking SFR estimators computed assuming a constant SFR over 1~Gyr  would lower the bias from $\sim25$\% to $\sim10$\% in the FUV band as the difference in the calibration coefficients is $\sim10$\%. In the U band the difference is even more important, the calibration coefficient differing by $\sim50$\%.
The second lesson, in link to the first one, is that the assumption of a constant SFR proves to be too simple. In the MIRAGE simulations there is a strong initial episode of star formation. This rapidly accumulates long--lived stars that contaminate star formation tracing bands for hundreds of Myr afterwards. Even if we have discarded this initial burst, it shows that strong episodes of star formation have lasting effects on the measure of the SFR. This exemplifies that variations of the SFR break the assumption of a constant SFR over long periods and that this has clear consequences. This calls for more realistic SFH to be considered or systematic biases may be left uncorrected. In other words, this recommendation to use SFR estimators calibrated on long timescales is valid for main sequence galaxies with long star formation episodes and not for starbursting galaxies for which the relative contamination will be much lower and for which calibration on short timescales are adequate \citep{oti2010a}. That is, we strongly recommend that SFR estimators are calibrated assuming a SFH consistent with the one expected for the sample under study, otherwise biases and errors will be introduced.

If random errors can be a problem for the measurement of the SFR in individual galaxies, over large samples they will average out. However such large samples will remain affected by systematic biases. There has been a long--standing, redshift--dependent, tension between the integral of the SFR density of the Universe corrected for the return fraction and the stellar mass density. In their review, \cite{madau2014a} found an excess of 0.2~dex for the integrated SFR density. Numerous explanations have been put forward involving biases on the measurements of either quantity or in the extrapolation of luminosity functions (see the aforementioned review). The excess we find in the present study would explain a part of the discrepancy though not necessarily the entirety. They adopt SFR estimators calibrated over 300~Myr which reduce the bias we find but do not eliminate it. However, the relatively smooth variations of the SFR over long periods in the MIRAGE simulations likely provide us with a lower bound of possible biases, with more violent variations yielding larger biases that cannot be corrected for by simply changing the assumption of the duration of the star formation episode. The use of numerical simulations such as the present ones may prove to be useful for determining new, statistically accurate SFR estimators across various redshifts and galaxy parameters. Such an approach was for instance adopted by \cite{wilkins2012b} using a semi--analytic model. Interestingly, contrary to what we find here they have determined that standard estimators from the FUV, NUV and \textit{u} bands provide unbiased estimates at $z=0$, although with an uncertainty of 20\% at $z=0$, and 30\% at $z=6$. However they find an evolution of the UV calibration coefficient with redshift, which may be due to the same processes that we have shown in the current study. Direct comparison between the two studies is difficult though. The fundamental difference is that they obtained the SFH of a large number of galaxies using the \textsc{galform} semi--analytic model \citep{cole2000a, baugh2005a} and computed average calibration factor at a handful of redshift whereas we followed a small sample of galaxies with their SFH determined over nearly $\sim700$~Myr through the means of state--of--the--art hydrodynamical simulations with highly refined baryonic physics.

\section{Conclusion\label{sec:conclusion}}

In this paper we have investigated the impact of the SFH properties on the measurement of the SFR in galaxies through classical estimators. To do so, we have used the state--of--the--art MIRAGE sample which consists of a set of 23 detailed hydrodynamical simulations of gas--rich mergers and isolated galaxies, including highly refined star formation prescriptions. It is designed to reproduce the characteristics of MASSIV \citep{contini2012a}, a spectroscopic sample of $1<z<2$ galaxies. Passed the initial relaxation, the SFH show little to no long term evolution but have rapid variations of the SFR with an amplitude of $\mathrm{\sigma_{SFR}\simeq0.34\pm0.06\times\left<SFR\right>}$. Peak--to--peak variations are of the order of a few. We have then combined the simulated SFH with the cigale SED modelling code (Boquien et al., Burgarella et al., in prep.) to compute the global spectrum for each simulation every Myr in an ideal case in order to isolate the impact of the variation of the SFR from any other potential effect: we took into account stellar populations and nebular emission free of dust obscuration while assuming a perfectly known IMF and no intrinsic uncertainties from the models.

Considering five different SFR estimators (Lyman continuum, FUV, NUV and U bands, and the TIR emission of buried galaxies), we have found that:
\begin{enumerate}
 \item On average all estimators except for the Lyman continuum overestimate the SFR from typically $\sim25$\% for SFR(FUV) to $\sim65$\% for SFR(U), when such estimators have been calibrated assuming a constant SFR over 100~Myr. This is chiefly due to the contribution of stars living longer than 100~Myr.
 \item Rapid variations of the SFR contribute to an increase of the uncertainty on the instantaneous SFR but have little long term effect on the measurement of the SFR.
 \item Slow variations of the SFR on timescales of a few tens of Myr and longer lead both to an increase of the noise on the measurement of the instantaneous SFR, and to systematic overestimates of the SFR for all tracers except for the Lyman continuum.
 \item The amplitudes of the systematic overestimates depend on the metallicity. For the same SFH, more metal--poor galaxies will be the most affected.
 \end{enumerate}


Finally, if we find that the Lyman continuum is the best tracer in the ideal case we have considered, it nevertheless suffers from specific uncertainties and observational difficulties. The widely used rest--frame FUV suffers for biases if applied na\"ively. These biases can be at least partially compensated for by using specially tailored SFR estimators for given redshifts and galaxy properties. At the very least, for main sequence galaxies we suggest adopting SFR estimators calibrated over 1~Gyr rather than the commonly used 100~Myr to take into account contamination by long--lived stars. Shorter timescales may still be considered for starburst galaxies or star--forming regions within galaxies for which the relative contamination from old stars should be lower.

\begin{acknowledgements}
We are particularly grateful to the referee for constructive and insightful comments that have helped clarify the article. We would like to thank Paola di Matteo for interesting discussions at the origin of this paper. MB would like to thank Robert Kennicutt and Michele Trenti for useful discussions that have helped improve this article.
\end{acknowledgements}

\appendix
\section{SFR estimators\label{sec:sfr-estimators}}
\begin{table*}
 \centering
 \begin{tabular}{cccccc}
  \hline\hline
$Z$  & Lyman continuum & FUV & NUV & U & TIR\\
     & log[M$_\odot$~yr$^{-1}$/(photons~s$^{-1}$)] & log[M$_\odot$~yr$^{-1}$/(W~Hz$^{-1}$)] & log[M$_\odot$~yr$^{-1}$/(W~Hz$^{-1}$)] & log[M$_\odot$~yr$^{-1}$/(W~Hz$^{-1}$)] & log[M$_\odot$~yr$^{-1}$/W]\\\hline
\multicolumn{6}{c}{10~Myr}\\\hline
0.0001 & $-53.427$ & $-20.923$ & $-20.818$ & $-20.705$ & $-36.564$\\
0.0004 & $-53.301$ & $-20.927$ & $-20.825$ & $-20.680$ & $-36.503$\\
0.004  & $-53.252$ & $-20.910$ & $-20.829$ & $-20.760$ & $-36.500$\\
0.008  & $-53.198$ & $-20.922$ & $-20.845$ & $-20.760$ & $-36.486$\\
0.02   & $-53.093$ & $-20.905$ & $-20.856$ & $-20.771$ & $-36.448$\\
0.05   & $-52.924$ & $-20.901$ & $-20.878$ & $-20.755$ & $-36.422$\\\hline
\multicolumn{6}{c}{50~Myr}\\\hline
0.0001 & $-53.439$ & $-21.139$ & $-21.031$ & $-20.897$ & $-36.697$\\
0.0004 & $-53.310$ & $-21.127$ & $-21.030$ & $-20.892$ & $-36.646$\\
0.004  & $-53.256$ & $-21.098$ & $-21.027$ & $-20.948$ & $-36.640$\\
0.008  & $-53.201$ & $-21.093$ & $-21.034$ & $-20.956$ & $-36.622$\\
0.02   & $-53.095$ & $-21.053$ & $-21.021$ & $-20.944$ & $-36.578$\\
0.05   & $-52.925$ & $-21.015$ & $-21.014$ & $-20.920$ & $-36.535$\\\hline
\multicolumn{6}{c}{100~Myr}\\\hline
0.0001 & $-53.439$ & $-21.200$ & $-21.101$ & $-20.990$ & $-36.744$\\
0.0004 & $-53.310$ & $-21.177$ & $-21.088$ & $-20.976$ & $-36.689$\\
0.004  & $-53.256$ & $-21.144$ & $-21.081$ & $-21.020$ & $-36.682$\\
0.008  & $-53.201$ & $-21.134$ & $-21.084$ & $-21.026$ & $-36.662$\\
0.02   & $-53.095$ & $-21.090$ & $-21.066$ & $-21.008$ & $-36.615$\\
0.05   & $-52.925$ & $-21.042$ & $-21.049$ & $-20.980$ & $-36.569$\\\hline
\multicolumn{6}{c}{500~Myr}\\\hline
0.0001 & $-53.439$ & $-21.279$ & $-21.206$ & $-21.169$ & $-36.827$\\
0.0004 & $-53.310$ & $-21.247$ & $-21.185$ & $-21.162$ & $-36.781$\\
0.004  & $-53.256$ & $-21.202$ & $-21.160$ & $-21.175$ & $-36.771$\\
0.008  & $-53.201$ & $-21.182$ & $-21.152$ & $-21.166$ & $-36.744$\\
0.02   & $-53.095$ & $-21.127$ & $-21.123$ & $-21.132$ & $-36.693$\\
0.05   & $-52.926$ & $-21.064$ & $-21.087$ & $-21.086$ & $-36.643$\\\hline
\multicolumn{6}{c}{1~Gyr}\\\hline
0.0001 & $-53.439$ & $-21.296$ & $-21.238$ & $-21.242$ & $-36.860$\\
0.0004 & $-53.311$ & $-21.261$ & $-21.211$ & $-21.232$ & $-36.817$\\
0.004  & $-53.256$ & $-21.209$ & $-21.176$ & $-21.228$ & $-36.807$\\
0.008  & $-53.202$ & $-21.186$ & $-21.164$ & $-21.211$ & $-36.778$\\
0.02   & $-53.095$ & $-21.128$ & $-21.130$ & $-21.168$ & $-36.727$\\
0.05   & $-52.926$ & $-21.064$ & $-21.090$ & $-21.115$ & $-36.681$\\\hline
 \end{tabular}
 \caption{Decimal logarithm of the calibration coefficient for the 5 star formation tracers considered in this study for a \cite{chabrier2003a} IMF between 0.1~M$_\odot$ and 100~M$_\odot$, each metallicity available in the \cite{bruzual2003a} stellar populations models, and for a constant SFR over 100~Myr and 1~Gyr. For reference, we have also included calibrations for a constant SFR over 10~Myr, 50~Myr, and 500~Myr. Calibrations on short timescales have been studied in detail \cite{oti2010a} for instance.}
\end{table*}

\bibliographystyle{aa}
\bibliography{article}

\end{document}